\documentstyle[11pt,Churchill,times]{article}
 
\newcommand{\etal}{et~al.\ }
\newcommand{\PVdblt}{{\rm P}\kern 0.1em{\sc v}~$\lambda\lambda 1117, 1128$}
\newcommand{\CIVdblt}{{\rm C}\kern 0.1em{\sc iv}~$\lambda\lambda 1548, 1550$}
\newcommand{\MgIIdblt}{{\rm Mg}\kern 0.1em{\sc ii}~$\lambda\lambda 2976, 2803$}
\newcommand{\NVdblt}{{\rm N}\kern 0.1em{\sc v}~$\lambda\lambda 1238, 1242$}  
\newcommand{\SVIdblt}{{\rm S}\kern 0.1em{\sc vi}~$\lambda\lambda 933, 944$} 
\newcommand{\OVIdblt}{{\rm O}\kern 0.1em{\sc vi}~$\lambda\lambda 1031, 1037$} 
\newcommand{\SiIIdblt}{{\rm Si}\kern 0.1em{\sc ii}~$\lambda\lambda 1190, 1193$} 
\newcommand{\SiIVdblt}{{\rm Si}\kern 0.1em{\sc iv}~$\lambda\lambda 1393, 1402$} 
\newcommand{\PV}{\hbox{{\rm P}\kern 0.1em{\sc v}}}
\newcommand{\AlI}{\hbox{{\rm Al}\kern 0.1em{\sc i}}}
\newcommand{\AlII}{\hbox{{\rm Al}\kern 0.1em{\sc ii}}}
\newcommand{\AlIII}{{\hbox{\rm Al}\kern 0.1em{\sc iii}}}
\newcommand{\CaII}{\hbox{{\rm Ca}\kern 0.1em{\sc ii}}}
\newcommand{\CII}{\hbox{{\rm C}\kern 0.1em{\sc ii}}}
\newcommand{\CIIe}{\hbox{{\rm C$^{\ast}$}\kern 0.1em{\sc ii}}}
\newcommand{\CIII}{\hbox{{\rm C}\kern 0.1em{\sc iii}}}
\newcommand{\CIV}{\hbox{{\rm C}\kern 0.1em{\sc iv}}}
\newcommand{\CV}{\hbox{{\rm C}\kern 0.1em{\sc v}}}
\newcommand{\HI}{\hbox{{\rm H}\kern 0.1em{\sc i}}}
\newcommand{\HII}{\hbox{{\rm H}\kern 0.1em{\sc ii}}}
\newcommand{\Lya}{\hbox{{\rm Ly}\kern 0.1em$\alpha$}}
\newcommand{\Lyb}{\hbox{{\rm Ly}\kern 0.1em$\beta$}}
\newcommand{\Lyg}{\hbox{{\rm Ly}\kern 0.1em$\gamma$}}
\newcommand{\Lyd}{\hbox{{\rm Ly}\kern 0.1em$\delta$}}
\newcommand{\Lye}{\hbox{{\rm Ly}\kern 0.1em$\epsilon$}}
\newcommand{\Lyphi}{\hbox{{\rm Ly}\kern 0.1em$\phi$}}
\newcommand{\Lyfive}{\hbox{{\rm Ly}\kern 0.1em$5$}}
\newcommand{\Lysix}{\hbox{{\rm Ly}\kern 0.1em$6$}}
\newcommand{\Lyseven}{\hbox{{\rm Ly}\kern 0.1em$7$}}
\newcommand{\Lyeight}{\hbox{{\rm Ly}\kern 0.1em$8$}}
\newcommand{\Lynine}{\hbox{{\rm Ly}\kern 0.1em$9$}}
\newcommand{\Lyten}{\hbox{{\rm Ly}\kern 0.1em$10$}}
\newcommand{\HeI}{\hbox{{\rm He}\kern 0.1em{\sc i}}}
\newcommand{\HeII}{\hbox{{\rm He}\kern 0.1em{\sc ii}}}
\newcommand{\FeI}{\hbox{{\rm Fe}\kern 0.1em{\sc i}}}
\newcommand{\FeII}{\hbox{{\rm Fe}\kern 0.1em{\sc ii}}}
\newcommand{\FeIII}{\hbox{{\rm Fe}\kern 0.1em{\sc iii}}}
\newcommand{\MnII}{\hbox{{\rm Mn}\kern 0.1em{\sc ii}}}
\newcommand{\MgI}{\hbox{{\rm Mg}\kern 0.1em{\sc i}}}
\newcommand{\MgII}{\hbox{{\rm Mg}\kern 0.1em{\sc ii}}}
\newcommand{\MgIII}{\hbox{{\rm Mg}\kern 0.1em{\sc iii}}}
\newcommand{\NI}{\hbox{{\rm N}\kern 0.1em{\sc i}}}
\newcommand{\NII}{\hbox{{\rm N}\kern 0.1em{\sc ii}}}
\newcommand{\NIII}{\hbox{{\rm N}\kern 0.1em{\sc iii}}}
\newcommand{\NV}{\hbox{{\rm N}\kern 0.1em{\sc v}}}
\newcommand{\OVI}{\hbox{{\rm O}\kern 0.1em{\sc vi}}}
\newcommand{\OI}{\hbox{{\rm O}\kern 0.1em{\sc i}}}
\newcommand{\OII}{\hbox{[{\rm O}\kern 0.1em{\sc ii}]}}
\newcommand{\OIV}{\hbox{{\rm O}\kern 0.1em{\sc iv}]}}
\newcommand{\SVI}{{\rm S}\kern 0.1em{\sc vi}}
\newcommand{\SiI}{\hbox{{\rm Si}\kern 0.1em{\sc i}}}
\newcommand{\SiII}{\hbox{{\rm Si}\kern 0.1em{\sc ii}}}
\newcommand{\SiIII}{\hbox{{\rm Si}\kern 0.1em{\sc iii}}}
\newcommand{\SiIV}{\hbox{{\rm Si}\kern 0.1em{\sc iv}}}
\newcommand{\SII}{\hbox{{\rm S}\kern 0.1em{\sc ii}}}
\newcommand{\SIII}{\hbox{{\rm S}\kern 0.1em{\sc iii}}}
\newcommand{\NaI}{\hbox{{\rm Na}\kern 0.1em{\sc i}}}
\newcommand{\kms}{\hbox{km~s$^{-1}$}}
\newcommand{\cmsq}{\hbox{cm$^{-2}$}}

\tighten
\begin{document}
 
 
\lefthead{CHURCHILL ET AL.}
\righthead{MINI--BAL QUASAR AT $z=4.59$}

 \submitted{The Astronomical Journal, {\it in press}}
\title{An Unusual Mini--BAL Quasar at $\lowercase{z}=4.59$\altaffilmark{1}}
 
\thispagestyle{empty}
 
\author{Christopher~W.~Churchill$^{2}$, 
Donald P. Schneider$^{2}$, \\ Maarten Schmidt$^{3}$, and
James E. Gunn$^{4}$}

\affil{$^{2}$ Department of Astronomy and Astrophysics, 
       Pennsylvania State University, 
       University Park, PA 16802 \\
       $^{3}$ California Institute of Technology, 
       Pasadena, CA 91125 \\
       $^{4}$ Princeton University Observatory, 
       Princeton, NJ 08544 }

\altaffiltext{1}{Based in part on observations obtained at the
W.~M. Keck Observatory, which is jointly operated by the University of
California and the California Institute of Technology.}

\begin{abstract}
The $z = 4.591$ quasar PC~1415+3408 exhibits very strong associated
metal--line absorption from the {\NVdblt}, {\SiIVdblt}, and {\CIVdblt}
doublets spanning the velocity interval $-1700 \leq v \leq 0$~{\kms}.
Also present, are detached absorption troughs in {\NV} and {\CIV}
spanning $-5000 \leq v \leq -3000$~{\kms}; this is characteristic of
broad absorption line (BAL) quasars, but the small overall velocity
spread suggests that PC~1415+3408 be classified as a ``Mini--BAL''
quasar. 
The {\NV} doublet is consistent with black saturation over
the velocity interval $-1200$ to $-500$~{\kms}; black {\NV} absorption
is extraordinary in all classes of quasars at all redshifts.
Over this velocity interval, the {\CIV} doublet is severely blended,
but also consistent with black saturation.
The material over this range of velocity appears to fully occult the
continuum source, the broad emission line region, and any material
that could give rise to scattered light.
In view of a unified scenario for BAL and Mini--BAL absorption, these
facts imply that the quasar is being viewed along a preferred
direction.
On the other hand, the black Mini--BALs in PC~1415+3408 could be
explained if the BAL flow has an unusual geometry compared to the
population of BAL quasars, and/or the spatial extent of a scattering
region is small at the lower velocities ($-1700 \leq v \leq
0$~{\kms}).
\end{abstract}

\section{Introduction}
\label{sec:intro}
 
The absorption lines from metals with $z_{\rm abs} \simeq
z_{\rm em}$ are useful probes of the kinematics and chemical and
ionization conditions of the material surrounding quasars, whether the
material gives rise to broad absorption lines (BALs) or narrower
absorption lines
(e.g.\ \cite{weymann78}; \cite{raybaby81}; \cite{briggs84};
\cite{turnshek84}; \cite{morris}; \cite{anderson}; \cite{turnshek88};
\cite{weymann91}; \cite{korista93}; \cite{turnshek95}; \cite{arav96};
\cite{turnshek96}; \cite{turnshek97a}, 1997b\nocite{turnshek97b}; 
\cite{freddie}; \cite{hamannbj97},
1997b\nocite{hamann97},
1997c\nocite{freddieetal97}; \cite{barlow}; \cite{arav98}).
These absorption lines sample environments in which the chemical and
ionization conditions have presumably been influenced by the
central engine of the quasar active nucleus (see Arav, Shlosman, \&
Weymann 1997\nocite{AGNconf}).

BAL gas is thought to be material within $\sim 1$ kpc of the quasar
that is undergoing a high velocity outflow, typically $-5000$ to
$-25,000$~{\kms} (\cite{turnshek88}; \cite{weymann91}).
BAL quasars comprise roughly 10\% of all quasars (however, see
\cite{goodrich97}; \cite{krolik98}), are radio--quiet
(\cite{stocke92}; however, see \cite{becker97}), and are quiet, or
self--absorbed, in soft X--rays ($h\nu < 2$~keV: \cite{kopko94};
\cite{green96}) and hard X--rays ($2 \leq h\nu \leq 10$~keV:
\cite{gallagher}).
Though there are many interpretations of these trends, a unified
picture has been proposed in which all quasars have BAL flows and that
various viewing angles through the outflowing material result in
different observed absorption properties.
Most unified pictures, as suggested by polarization studies (e.g.\
\cite{cohen95}), have the BAL material {\it originating\/} from a
disk geometry (e.g.\ \cite{emmering92}; \cite{dekool95};
\cite{murray95}).

\begin{figure*}[t]
\plotfiddle{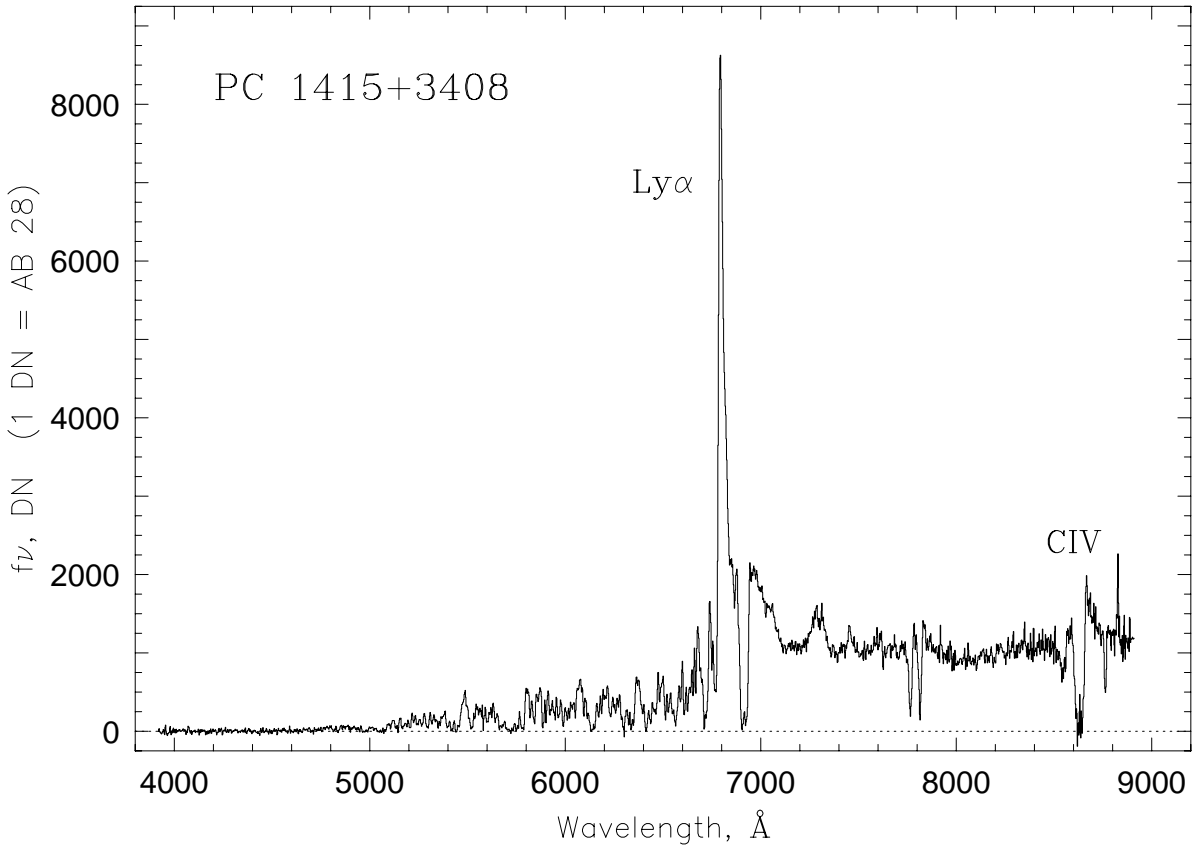}{4.25in}{0.}{65.}{65.}{-257}{-40}
\protect\caption
{\footnotesize
LRIS/Keck spectrum of PC~1415+3408 obtained with the 300 lines
mm$^{-1}$ grating. The resolution of the data is $R \simeq 900$ (in
0.7{\arcsec} seeing using a 1.0{\arcsec} slit) with
2.6~{\AA}~pix$^{-1}$. The unit of flux along the vertical axis is 
$2.3\times 10^{-31}$ ergs {\cmsq} s$^{-1}$ Hz$^{-1}$ ($ = $ AB
magnitude 28).  \label{fig:fig1}
}
\end{figure*}

An apparently rare class of so--called ``Mini--BAL'' quasars have been
identified (\cite{turnshek88}; \cite{barlowAGN}).
These quasars often exhibit flat--bottomed (but not necessarily black
bottomed) absorption profiles with overall velocity spreads less than
2000~{\kms}.
The relationship between Mini--BAL and BAL quasars is not yet
understood.
From the point of view of ultraviolet rest--frame (UV) spectra,
Mini--BAL quasars may prove to be the most useful for deducing the
chemical and ionization conditions of material involved in BAL flows
(\cite{aravAGN}), since they do not suffer the extreme blending seen
in BAL quasars.
Ultimately, however, any working model of BAL flows will need to
survive observational probings not only with UV rest--frame data, but
with radio and both soft and hard X--ray data as well.

In addition to multi--band observations, studies are needed that
extend the sample of BAL and Mini--BAL quasars to the highest redshifts.
Nearly 100 quasars at $z\geq4.0$ have now been discovered and the
fraction with BAL features appears to be the same as at lower
redshifts (e.g.\ \cite{apmsurvey}).
Extending BAL quasar studies to $z>4$ would yield a chronological
baseline on par with studies of the {\Lya} absorbers for firmly
establishing cosmic evolution (or lack of) in a class of astronomical
object.
Unlike (but complementary to) studies of {\Lya} absorption,
investigations of BAL quasars probe the most energetic and active
sites in the Universe. 

In this paper, we present high--quality spectra of one of the few
known $z \geq 4.5$ quasars, PC~1415+3408, which possesses
very unusual mini--BAL spectral characteristics.
PC~1415+3408 was first reported by Schneider, Schmidt, \& Gunn
(1997\nocite{ssg97}, hereafter SSG97) and was observed to have 
peculiar emission and/or absorption features as seen in a
low--resolution ($\sim 25$~{\AA}) and low signal--to--noise spectrum.
Furthermore, an accurate emission redshift measurement was difficult;
they tentatively assigned $z_{\rm em} = 4.6$, but noted that the
complex {\CIV} emission profile could be placed at $z_{\rm em} =
4.76$. 
We present a robust emission redshift and investigate the emission and
absorption line properties of PC~1415+3408.
Throughout this work we assume $H_{0}=50$~{\kms}~Mpc$^{-1}$, $q_{0} =
0.5$.
We do not make a distinction between the categorical terms
``associated'' and ``intrinsic'' absorption for describing material
with $z_{\rm abs} \simeq z_{\rm em}$ (for a clear exposition see
\cite{barlowAGN}).

\section{Spectroscopic Observations}
\label{sec:data}

\begin{figure*}[t]
\plotfiddle{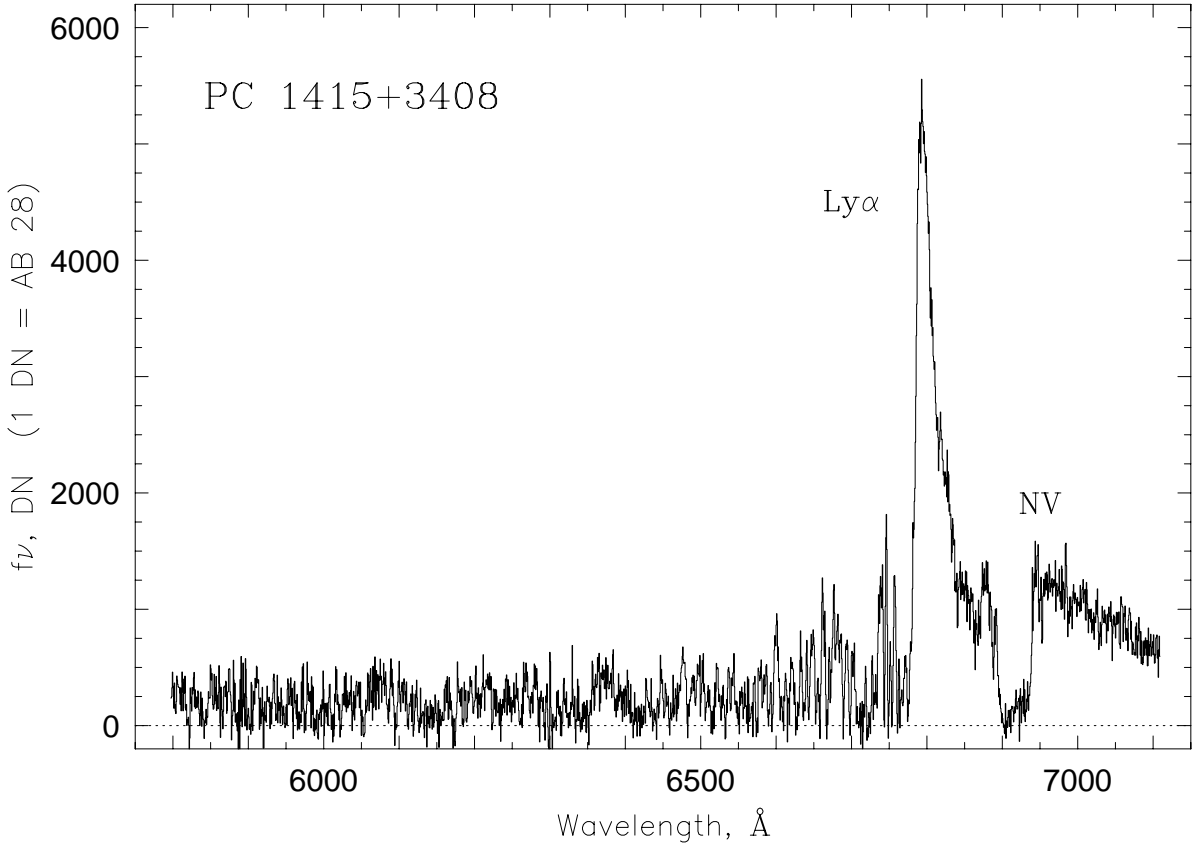}{4.25in}{0.}{65.}{65.}{-257}{-40}
\protect\caption
{\footnotesize
LRIS/Keck spectrum of PC~1415+3408 obtained with the 1200 lines
mm$^{-1}$ grating. The resolution of the data is $R \simeq 4000$ with
0.65~{\AA}~pix$^{-1}$.  \label{fig:fig2}
}
\end{figure*}

Spectra of PC~1415+3408 were acquired with the Low Resolution
Imaging Spectrograph (LRIS) on the Keck~I Telescope (\cite{oke95}).
An 1800 second exposure was obtained on 1996 April 14, using
the 1200 line~mm$^{-1}$ grating and a 0.7{\arcsec} slit under   
$\sim 1${\arcsec} seeing conditions.
The resolution of the spectrum is $R\sim4000$ with
$\sim0.64$~{\AA}~pix$^{-1}$, and the wavelength coverage is
5797.6--7108.8~{\AA}.
An additional 1800 second exposure using the 300 line~mm$^{-1}$
grating with a 1.0{\arcsec} slit was obtained on 1997 April 3.
The seeing was $\sim 0.7${\arcsec}, resulting in a resolution of
$R\sim900$ with $\sim 2.4$~{\AA}~pix$^{-1}$.  
The wavelength coverage is 3917.8--8908.9~{\AA}.
For all observations, the spectrograph slit was set at the parallactic
angle to minimize differential refraction.
The spatial scale on the CCD is 0.213{\arcsec}~pix$^{-1}$.

The raw data frames were processed in the standard fashion, including
bias subtraction and flat fielding.  
The spectra were extracted using the optimal algorithm of Horne
(1986\nocite{horne}).
The wavelength scale was set (for both gratings) with cubic fits to
lines from Hg, Kr, and Ar discharge lamps.  
The resulting RMS errors are 0.09~{\AA} for the 1200~lines~mm$^{-1}$
grating and 0.23~{\AA} for the 300~lines~mm$^{-1}$ grating.
For the latter observations, the seeing was less than the slit width,
resulting in a $\sim 1$~pixel uncertainty in the wavelength scale zero
point.
The flux calibration and the removal of atmospheric absorption bands
were performed using the flux standards of Oke \& Gunn
(1983\nocite{oke-gunn}).
The fully reduced and calibrated spectra are shown in
Figure~\ref{fig:fig1} (300~lines~mm$^{-1}$) and Figure~\ref{fig:fig2}
(1200~lines~mm$^{-1}$).

\section{PC 1415+3408}
\label{sec:properties}

\subsection{UV Rest--Frame Properties}

\begin{figure*}[t]
\plotfiddle{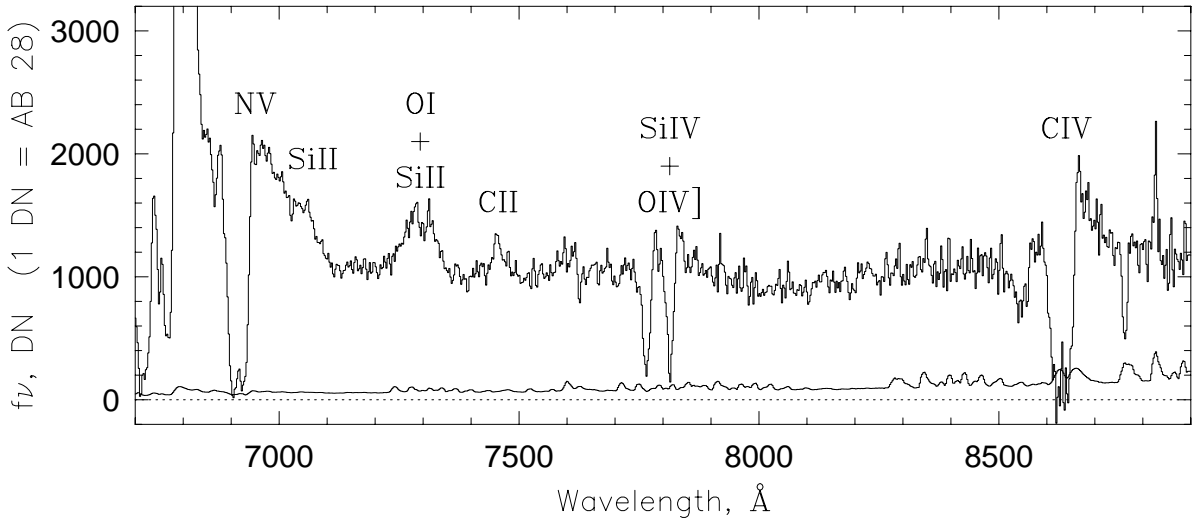}{2.75in}{0.}{65.}{65.}{-257}{-100}
\protect\caption
{\footnotesize
The 300~lines~mm$^{-1}$ spectrum of PC~1415+3408 redward of {\Lya}
emission.  The broad emission lines are labeled.
The narrow ``absorption'' line at 8762~{\AA} and the narrow
``emission'' lines at 8827~{\AA} and on the {\CIV} broad emission line
at 8670~{\AA} are artifacts of the sky subtraction.
The $1\sigma$ error array is given by the spectrum just above the zero
point. \label{fig:fig3}
}
\end{figure*}

In Figure~\ref{fig:fig3} we show the 300~lines~mm$^{-1}$ spectrum
redward of the {\Lya} emission line with the vertical scale set to
emphasize the continuum features.
PC~1415+3408 has strong, broad high--ionization emission lines from 
{\NV} $\lambda 1240.1$, 
{\SiIV}+{\OIV} $\lambda 1400.0$, 
and {\CIV} $\lambda 1549.1$.
The cores and blue wings of these emission lines are strongly
absorbed.
In absorption, the {\NVdblt} doublet members are partially resolved,
the {\SiIVdblt} doublet members are well resolved, and the {\CIVdblt}
doublet members are fully blended.
There are also weaker, low--ionization emission lines from {\SiII}
$\lambda 1263.0$, {\OI}+{\SiII} $\lambda 1304.5$, and {\CII} $\lambda
1334.5$.
The {\SiII} emission is on the red wing of the broad {\NV} emission
line, making it difficult to accurately determine the {\SiII}
strength.
The {\OI}+{\SiII} feature appears to be among the strongest seen in
quasars (of all redshifts) reported in the literature and has a weak
absorption feature that is slightly redshifted.
As seen in Figure~\ref{fig:fig1}, there is a clear Lyman limit break
at 5080~{\AA}, which corresponds to $z=4.575$.
The lack of a recovery by $704$~{\AA} in the rest--frame implies
$N({\HI}) > 10^{18.2}$~{\cmsq}.
There is strong {\Lya} absorption at 6777~{\AA} that corresponds to
the break (see Figures~\ref{fig:fig1} and \ref{fig:fig2}).
In Table~\ref{tab:properties}, we list the overall properties of
PC~1415+3408.
The $g_4$ and $r_4$ filters each have widths of approximately
$450$~{\AA} and are centered at $4930$ and $6550$~{\AA}, respectively.
The quantity $AB_{1450}$ is the observed $AB$ magnitude, corrected for
Galactic reddening, at $\lambda = 1450 (1+z)$, where $AB = -2.5\log f_{\nu}
- 48.60$ (see \cite{ssg89}).


The spectral energy index, $\alpha$, was determined by fitting the
function $f_{\nu} \propto \nu ^{\alpha}$ to interactively selected
continuum points redward of $\sim 6780$~{\AA} using the
300~lines~mm$^{-1}$ spectrum.
A possible observational systematic error in the slope could be
differential refraction, though we note that the seeing was better
than the projected width of the slit and we aligned the slit with the
parallactic angle.
To estimate possible systematics in the fitting procedure itself, we
performed multiple interactive fits.
Just over 65\%  
of the total available wavelength coverage redward of the blended
{\Lya}+{\NV} emission line was employed.
This exercise yielded $\alpha = -0.45\pm0.05$.
Our $\alpha $ value is significantly different than that measured by
SSG97 ($\alpha = -1.2$) using lower quality data.
Normalization of the power law gives $AB_{1450} = 20.6$, from which
an absolute blue--band magnitude of $M_{B} = -25.9$ was
calculated following Schneider \etal (1989)\footnote{The zero--point
offset applied here is corrected to be half that applied in their work
(see \cite{ssg95}).}.
These values are in good agreement with those calculated by SSG97.

\begin{figure*}[b]
\plotfiddle{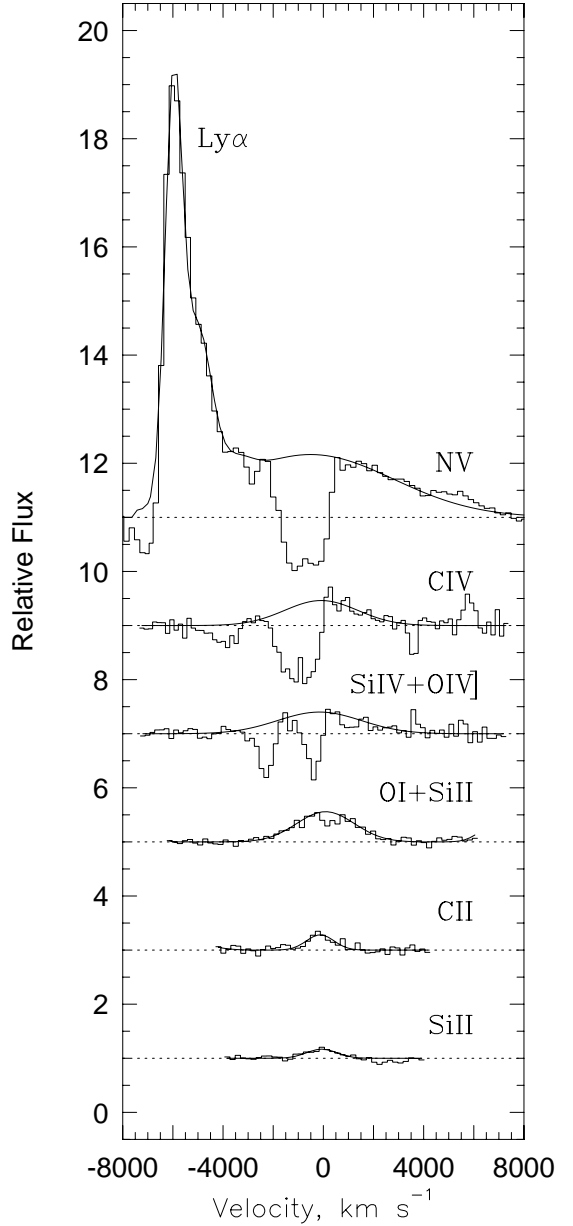}{0.in}{0.}{70.}{70.}{263}{72}
\parbox{3.5in}{\phantom{dummy}} \hfill
\parbox{3.5in}{\protect\caption{\vglue -2.25in 
\footnotesize
{\sc Fig}.~~4--- The emission lines and the {\NV}, {\CIV}, and {\SiIV}
absorption lines
of PC~1415+3408 as measured in the 300~lines~mm$^{-1}$ grating
spectrum.  For presentation, the spectra have been normalized by the
``source'' continuum, are offset vertically by a factor of two (the
respective zero points are 0, 2, 4, 6, 8, and 10), and are aligned in
velocity referenced to the quasar emission redshift.  \label{fig:fig4}
}}
\end{figure*}


Measurements of the emission line properties were obtained by 
interactively fitting a smooth third--order Legendre polynomial
to the ``source'' continuum redward of 6770~{\AA}, excluding emission
and absorption lines.
The emission lines were modeled with single component Gaussian
fits using a Levenberg--Marquardt least square minimizer
(\cite{more78}).
However, the {\Lya}+{\NV} blend required three Gaussian components.
Each component fit yielded a central wavelength, width, equivalent
width, and $1\sigma$ uncertainties in these quantities.
The resulting emission line profiles are shown in
Figure~\ref{fig:fig4}, and their measured properties are listed in 
Table~\ref{tab:emission}.
A redshift of $z_{\rm em} = 4.591\pm0.001$ was determined from the
weighted mean of the emission line central wavelengths (corrected to
vacuum).  
The weight for each line was taken to be $\sigma_{\lambda}^{-2}$,
where $\sigma _{\lambda}$ is the $1\sigma$ uncertainty in the central
wavelength.
The blended {\Lya}+{\NV} emission line was excluded.

The quantity $\Delta {\Lya}$, given by $[\lambda_{\small
\Lya}/(1+z_{\rm em}) - 1215.67]$, where $\lambda_{\small \Lya}$ is the
vacuum corrected wavelength of the {\Lya} emission peak, quantifies
the degree of blueward absorption in the rest--frame of the {\Lya}
emission line.
The offset measured for PC~1415+3408,  $\Delta {\Lya} = 0.8$~{\AA},
is the smallest measured at $z \geq 4$ as compared to the sample of 33
quasars (of which 13 were  $z \geq 4$) studied by Schneider, Schmidt,
\& Gunn (1991\nocite{ssg91}) [also see \cite{sbs88}]. 
This small $\Delta {\Lya}$ implies either that the {\Lya} emission
line is intrinsically very narrow or suffers absorption along both
wings ({\Lya} to the blue and high velocity, asymmetric, {\NV} to the
red) that conspire to leave its peak relatively unshifted.


For the highest redshift quasars, absorption blueward of {\Lya} emission
is very strong.
The ``flux deficit'' (\cite{oke82}), given by
\begin{equation}
D = \left< 1 - \frac{f_{\nu}({\rm observed})}
                    {f_{\nu}({\rm continuum})}  \right>,
\end{equation}
provides a quantitative measure of this absorption, where
$f_{\nu}({\rm observed})$ is the observed flux and $f_{\nu}({\rm
continuum})$ is the predicted unattenuated flux based upon
extrapolation of the spectral energy index.
This deficit is measure over two regimes.
The first, $D_{A}$, is measured from 1050 to 1170~{\AA} in the
rest--frame of the quasar and gives the mean absorption blueward
of the {\Lya} emission line and redward of the {\Lyb}+{\OVI} emission
line; it is designed to measure {\Lya}--only absorption.
The second, $D_{B}$, is measured from 920 to 1015~{\AA}
in the rest--frame of the quasar and gives the mean absorption
blueward of the {\Lyb}+{\OVI} emission line and redward of the Lyman
limit of the quasar.

The dominant source of error in the measurement of the flux decrement
is the uncertainty in the power--law continuum fit.
To estimate the uncertainty in our measurement, we used the $1\sigma$
spread in $\alpha$ and in $AB_{1450}$ that conspire to give the
extreme upper and lower values of $D$.
For PC~1415+3408, we measured $D_{A} \simeq 0.7$, with an uncertainty
of $0.06$.
This value is typical of $z \simeq 4.5$ quasars; $D_{A}$ is seen to
rise from $\sim 50$\%  at $z=4$ to $\sim 80$\% at $z\sim 5$
(\cite{ssg91}).
This has been shown to be consistent with an increase in the numbers
of {\Lya} clouds at higher redshifts
(\cite{giallongo}; \cite{jenkins91}; \cite{ssg91}).
We find $D_{B} \simeq 0.8$, with uncertainty $0.1$.
This value is slightly high compared to the mean value $0.65$ measured
for 13 $z>4$ non--BAL quasars (\cite{ssg91}), possibly due to 
strong associated metal lines (esp.\ {\OVI}, {\CIII}, {\NIII}).

In the 300~lines~mm$^{-1}$ spectrum (see Figure~\ref{fig:fig1}), there
is strong, broad absorption at 6143~{\AA}, which has the appearance of
a damped {\Lya} absorber (DLA) at $z=4.054$. 
If this is a DLA, the rest--frame equivalent width for {\Lya} is
$14.5\pm0.6$~{\AA}, which corresponds to $\log N({\HI}) \sim
21.0$~{\cmsq}.
However, in the 1200~lines~mm$^{-1}$ spectrum, the absorption appears
to break into a series of smaller, blended {\Lya} clouds, as seen in
Figure~\ref{fig:fig2}. 
This serves as a warning that DLA candidates at $z\sim 4$ in
spectra with 5--6~{\AA} resolution require follow--up verification at
higher resolutions.

\subsection{Radio and X--ray Properties}

\begin{deluxetable}{lc}
\tablenum{1}
\tablecolumns{2}
\tablewidth{275pt}
\tablecaption{Properties of PC~$1415+3408$}
\tablehead
{
\colhead{Parameter} & \colhead{Value} 
}
\startdata
Right ascension (1950.0) \dotfill & $ 14^{\rm h}$ $15^{\rm m}$ $46.6^{\rm s}$  \\
Declination (1950.0)     \dotfill & $+34^{\circ}$ $08^{\prime}$ $30\arcsec$     \\
$z_{\rm em}$             \dotfill & $4.591\pm 0.001$   \\
$g_{4}$                  \dotfill & $ 23.76 \pm0.17$   \\
$r_{4}$                  \dotfill & $ 21.37 \pm0.08$   \\
$\alpha$                 \dotfill & $ -0.45 \pm0.05$ \\
$AB_{1450}$              \dotfill & $ 20.57  \pm0.09 $ \\   
$M_{1450}$               \dotfill & $-26.3$           \\  
$M_{B}$                  \dotfill & $-25.9$           \\         
$D_{A}$                  \dotfill & $  0.70 \pm 0.06$  \\
$D_{B}$                  \dotfill & $  0.8 \pm 0.1$  \\  
$\Delta{\Lya}$           \dotfill & $  0.81 \pm 0.02$~{\AA}  \\
\enddata
\label{tab:properties}
\end{deluxetable}

PC~1415+3408 is not detected in the FIRST survey to a flux limit
$f_{\nu} (1.4~{\rm GHz}) \leq 1.0$~mJy.
Assuming a spectral power index of $\alpha = -0.5$ for both the radio
and optical bands, we obtained 
$L_{\nu}({\rm 6 cm}) \leq 10^{25.2}$~W~Hz$^{-1}$ and 
$L_{\nu}(4400~{\rm {\AA}}) = 10^{24.0}$~W~Hz$^{-1}$ for an
isotropically radiating point source.
Using $L_{\nu}({\rm 6 cm}) \simeq 10^{25}$~W~Hz$^{-1}$ as the
dividing line between radio--loud and radio--quiet quasars
(\cite{kellerman89}), PC~1415+3408 is at most a radio--moderate
quasar.
Another indicator for radio--loudness, 
$R = L_{\nu}({\rm 6 cm})/L_{\nu}(4400~{\rm {\AA}})$, is less than
$\simeq 15$, which also suggests a radio--moderate quasar at most
(\cite{kellerman94}). 
If the radio spectrum is flat ($\alpha = 0$), the upper limit on the
radio luminosity decreases by 30\%.
PC~1415+3408 is clearly not a radio--loud quasar.
Even if no BAL absorption was seen in PC~1415+3408, it is
statistically likely to be radio quiet, given that only 5--10\% of
optically selected quasars are radio loud at high redshifts (e.g.\
\cite{schmidt95}, and references therein).

In the X--ray band, there are no serendipitous or pointed
observations of PC~1415+3408 by either {\it ROSAT\/} 
or {\it ASCA},
though an upper limit of 0.05~counts~s$^{-1}$ was 
obtained from the
{\it ROSAT Bright Source Catalog}.
Radio--quiet quasars have $\alpha _{\rm ox} = -1.57\pm0.15$
(\cite{greenetal95}), where 
$\alpha _{\rm ox} = 0.348 \log [L_{\nu}(2500~{\rm \AA})/L_{\nu}(2~{\rm
keV})]$.
Assuming this value of $\alpha _{\rm ox}$, an isotropically radiating
point source, an X--ray energy index of $\alpha _{E} = -1$, and a
Galactic hydrogen column density of $N({\HI}) = 1.3 \times
10^{20}$~{\cmsq} toward PC~1415+3408 (\cite{stark92}), we 
\pagebreak calculated 
an expected X--ray luminosity of $L_{\nu}(2~{\rm keV}) = 8.5 \times
10^{22}$~W~Hz$^{-1}$ using {\sc pimms} (\cite{mukai97}).
This is an order of magnitude below the upper limit on $L_{\nu}(2~{\rm
keV})$ from the {\it ROSAT Bright Source Catalog}.

\section{The Associated Broad Absorption Lines}
\label{sec:absorption}

As seen in Figures~\ref{fig:fig1}--\ref{fig:fig3}, strong, broad
absorption is present in the cores of the three high ionization
emission lines.
The velocity alignment of these absorption features is shown in
Figure~\ref{fig:fig4}.
Over the the velocity interval $-1700$ to $\sim 0$~{\kms}, the {\NV}
and {\CIV} doublets are blended, but the two members of the {\SiIV}
doublet are resolved.
Both {\NV} and {\CIV} are consistent with ``black'' saturation 
($\sim {\rm zero}$ flux) over the velocity range $-1200$ to
$-500$~{\kms}.
An additional component, or ``secondary detached trough'', is evident
at velocities from $-2200$ to $-3500$~{\kms} for {\NV} and from
$-3000$ to $-5000$~{\kms} for {\CIV}.
In broad absorption line (BAL) quasars, ``double--trough'' broad
absorption is not uncommon, appearing $\sim 20$\% of the time
(\cite{weymann91}; \cite{korista93}).

That the broad absorption lines in PC~1415+3408 have $z_{\rm abs} \simeq
z_{\rm em}$ is highly suggestive that they are physically associated
with the quasar, that they do not arise in intervening objects.
Furthermore, there are other features that suggest the lines are
associated with the quasar.
The lines are characteristic of BAL flow in that they (1) are smoothed
bottomed and do not exhibit the complex kinematic structure observed
in galaxy halos, (2) are comprised of a primary and a detached trough,
a common feature of BALs, and (3) are significantly broader than the
$\leq 600$~{\kms} velocity spreads of galactic halos.
Since the {\CIV} absorption extends to less than $5000$~{\kms} from
the emission redshift (\cite{weymann91}), PC~1415+3408 should be
classified as a so--called ``Mini--BAL'' quasar (e.g.\
\cite{turnshek88}; \cite{barlowAGN}).
Examples of Mini--BAL quasars are Q~0449--13 at $z_{\rm em} = 3.09$
(\cite{barlowAGN}) and the $z_{\rm em} = 1.98$ radio--loud quasar
PHL~1157+0128 (\cite{aldcroft97}).
The {\CIV} profile of the latter is virtually identical to that of
PC~1415+3408, except that the secondary trough is centered at
$-5100$~{\kms} and has a spread of $\simeq 1900$~{\kms}.

\begin{deluxetable}{lcccccc}
\tablenum{2}
\tablecolumns{7}
\tablewidth{0pc}
\tablecaption{Emission Line Properties of PC~$1415+3408$}
\tablehead
{
\colhead{Tran} &
\colhead{$\lambda _{r}$, {\AA}} &
\colhead{$\lambda _{o}$, {\AA}} &
\colhead{$z_{\rm em}$}   &
\colhead{$W_{o}$, {\AA}} &
\colhead{$W_{r}$, {\AA}} &
\colhead{$\sigma_{r}$, {\AA}} 
}
\startdata
${\Lya}+{\NV}$\tablenotemark{a}    \dotfill & $1215.7+1240.1$ & $6792.4\pm 0.1$ 
& $4.589$  & $448\pm45$ & $80.1\pm 8.0$ & $1.23\pm 0.02$ \\
{\SiII}           \dotfill & $1263.0       $ & $7060.5\pm 0.7$ & $4.592\pm 0.001
$ & $  8.1\pm 0.4$ & $ 1.45\pm 0.07$ & $2.83\pm 0.09$ \\ 
${\OI}+{\SiII}$   \dotfill & $1304.5       $ & $7294.3\pm 1.4$ & $4.593\pm 0.001
$ & $ 39.3\pm 3.4$ & $ 7.03\pm 0.60$ & $5.00\pm 0.27$ \\
{\CII}            \dotfill & $1334.5       $ & $7456.3\pm 1.7$ & $4.589\pm 0.001
$ & $  9.8\pm 1.5$ & $ 1.76\pm 0.27$ & $2.48\pm 0.30$ \\
${\SiIV}+{\OIV}$  \dotfill & $1400.0       $ & $7814.4\pm 4.3$ & $4.583\pm 0.003
$ & $ 41.1\pm 5.8$ & $ 7.35\pm 1.04$ & $7.67\pm 0.73$ \\
{\CIV}            \dotfill & $1549.1       $ & $8657.5\pm 6.9$ & $4.590\pm 0.004
$ & $ 45.6\pm11.9$ & $ 8.15\pm 2.12$ & $7.95\pm 1.06$ \\
\enddata
\tablenotetext{a}{The central wavelength quoted is that of the narrow
peak of the {\Lya} line.}
\label{tab:emission}
\end{deluxetable}

\begin{figure*}[b]
\plotfiddle{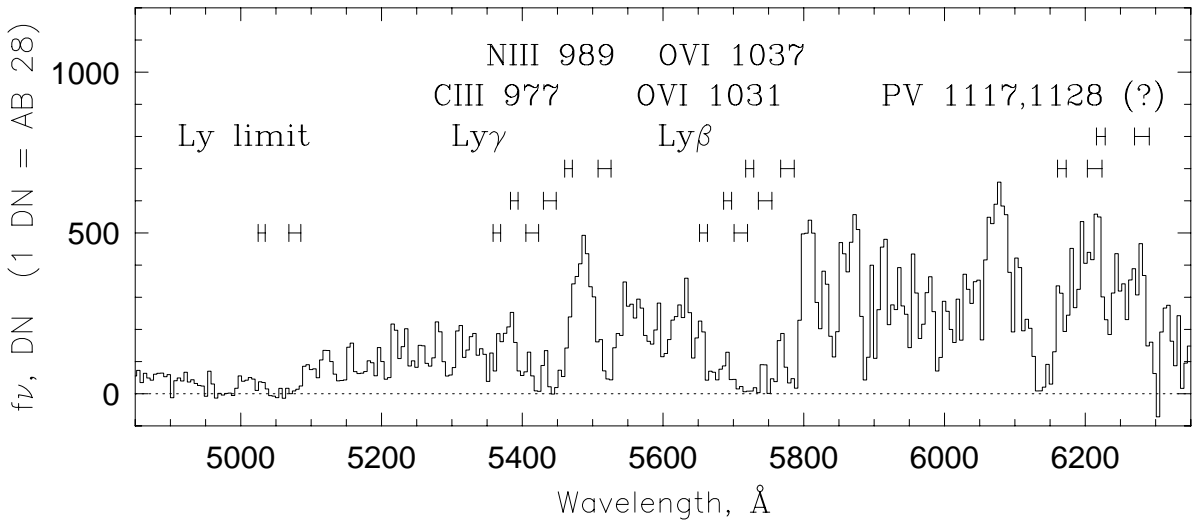}{2.55in}{0.}{65.}{65.}{-257}{-100}
\protect\caption
{\footnotesize
The 300~lines~mm$^{-1}$ spectrum of PC~1415+3408 blueward of {\Lya}
emission.  A significant portion of the absorption amongst the {\Lya}
forest is due to associated absorption from {\Lyg}+{\CIII}, {\NIII},
and {\Lyb}+{\OVI}.  The labeled bar ticks give the expected wavelength
regions of these species based upon the velocity spreads of {\Lya} and
{\NV}. \label{fig:fig5}
}
\end{figure*}

In Figure~\ref{fig:fig5}, we show the 4900--6100~{\AA} region
of the 300~lines~mm$^{-1}$ spectrum.
Bar ticks mark the expected wavelength {\it regions\/} of {\Lyg},
{\CIII} $\lambda 977$, {\NIII} $\lambda 989$, {\Lyb}, and {\OVIdblt},
based upon the $-5000 \leq v \leq -3000 $ and the $-1700 \leq v \leq
0$~{\kms} absorption troughs of {\CIV}.
Higher velocity absorption is evident in the {\Lyb} and {\OVI}
blend spanning  5700 to 5800~{\AA}; the blue wing is consistent with
the detached trough of the {\CIV}.
The {\SVIdblt} doublet may be present, but heavily blended by the
Lyman series.
The presence of {\PVdblt}, sometimes seen in BAL quasars, is
ambiguous.

\subsection{Exploring the Kinematics} 

\begin{figure*}[b]
\plotfiddle{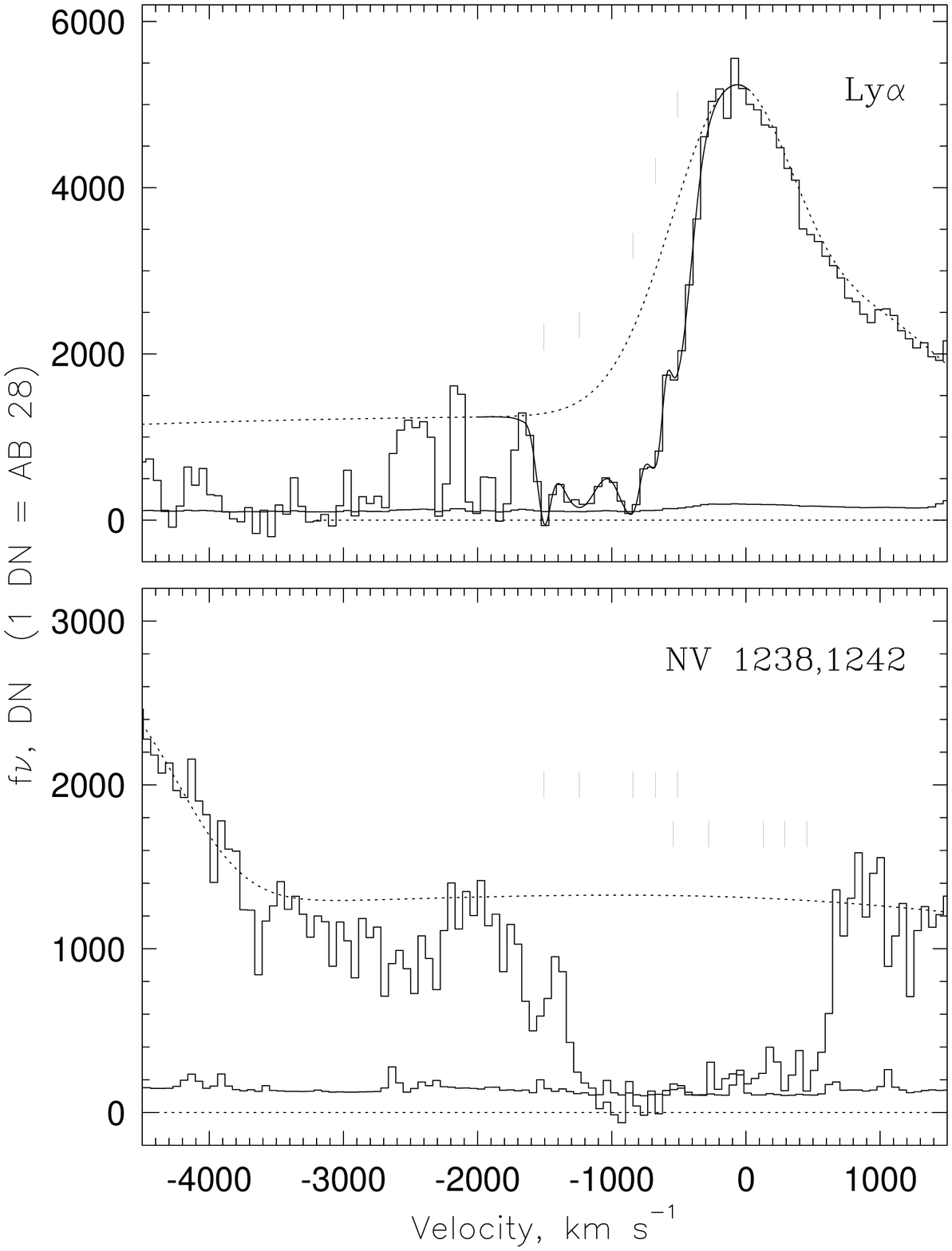}{0.in}{0.}{50.}{50.}{253}{152}
\parbox{3.5in}{\phantom{dummy}} \hfill
\parbox{3.5in}{\protect\caption{\vglue -3.1in 
\footnotesize
{\sc Fig}.~~6---
The associated {\Lya} (upper panel) and {\NVdblt} (lower panel)
absorption of PC 1415+3408 as measured in the 1200~lines~mm$^{-1}$
grating spectrum.  
The velocity zero--point is the quasar redshift, $z_{\rm em} = 4.591$.
The histogram data are the observed flux values and their
uncertainties.
Five ``components'' are shown based upon blended Gaussian fitting.
Ticks above the continuum give the Gaussian centroids.
In order of increasing outflow velocity, the absorption redshifts are 
$z = 4.5815$, $4.5785$, $4.5753$, $4.5678$, and $4.5629$.
The dot--dot lines are rough estimates of the continua and the solid
line shows the five--component model. \label{fig:fig6}
}}
\end{figure*}

For the remainder of this section, we focus on the primary absorption,
that in the velocity range $-1700 \leq v \leq 0$~{\kms}.
In Figure~\ref{fig:fig6}, we show the absorption blueward of the
{\Lya} emission line (upper panel) and of the blended {\NV} doublet
(lower panel), as measured in the 1200~line~mm$^{-1}$ spectrum.
The data are aligned in the rest--frame velocity of the quasar.
The quasar continuum blueward of {\Lya} emission was estimated 
by extrapolating a two--component Gaussian fit to the {\Lya}+{\NV}
blend with an underlying power law using the fiducial spectral energy
index, $\alpha = -0.5$ (since our value of $-0.45$ is consistent with
this fiducial value).
The resulting continuum is shown as a dot--dot curve in
Figure~\ref{fig:fig6}.

A series of Gaussians was fit to the flux values in the {\Lya}
absorption just blueward {\Lya} emission.
The free parameters for each Gaussian are the central wavelength,
amplitude, and width (unless the width of the component,
$\sigma_{c}$, is less than the instrumental resolution; in this case
$\sigma_{c}$ is set to $\lambda _{c}/(2.35R)$, where $\lambda _{c}$ is
the component centroid and $R=4000$ is the spectrum resolution).
We started with a two--component fit and increased the number of
components until a standard F--test yielded no further significant
gain (98\% confidence level) in the ``goodness'' of the fit, as
measured by the reduced $\chi^{2}$.
Our adopted fit has five Gaussian components.
Although these five {\Lya} components are not to be taken literally as
distinct absorbers, we emphasize the excellent alignment in
velocity with the blended {\NV} doublet.
In the bottom panel of Figure~\ref{fig:fig6}, we transpose the ticks
from the {\Lya} decomposition to the {\NV} doublet (upper ticks denote
the $\lambda 1238$ transition and lower ticks denote the $\lambda
1242$ transition); this alignment is strongly suggestive that the
{\Lya} absorption is metal--rich and gives rise to the {\NV}.

\begin{figure*}[b]
\plotfiddle{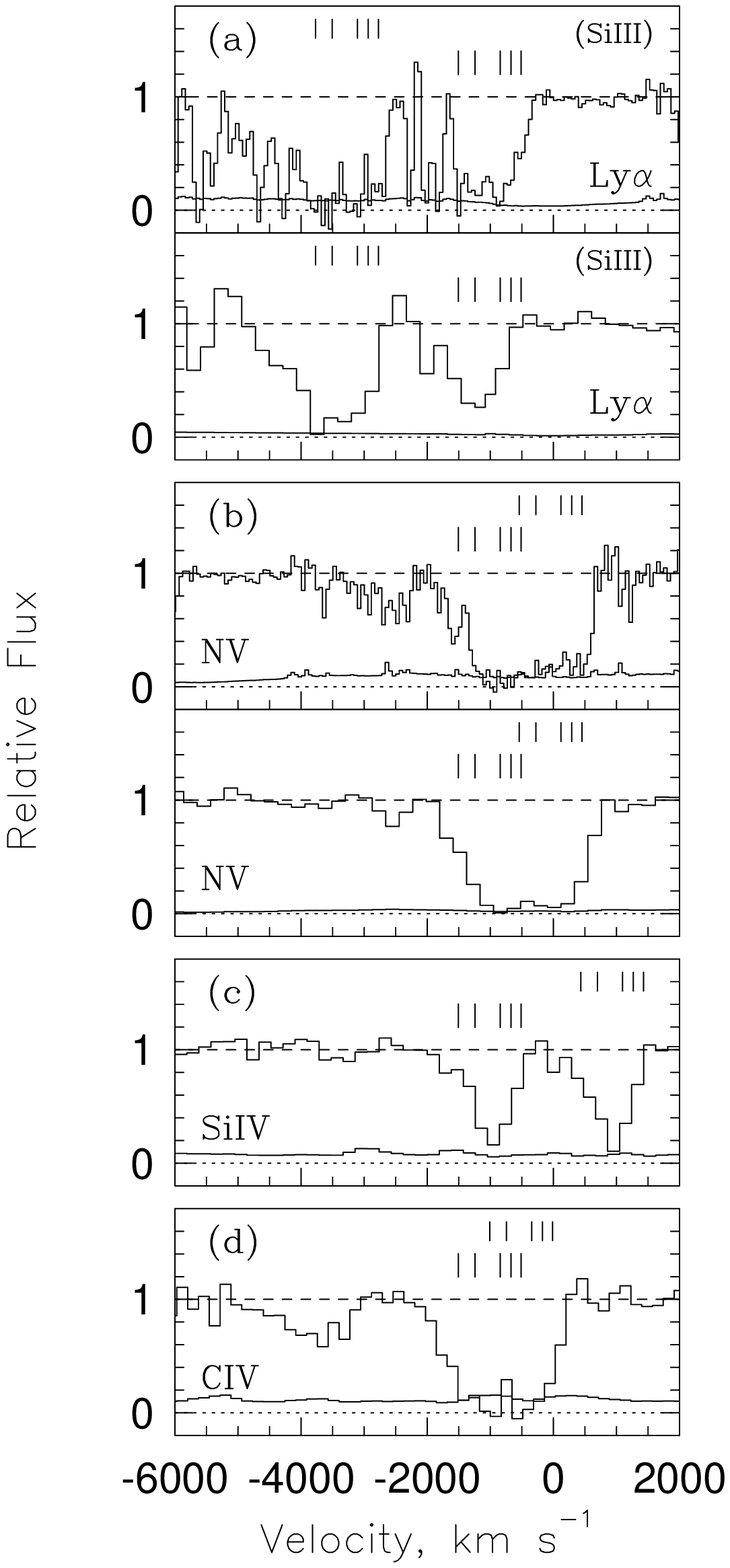}{0.in}{0.}{70.}{70.}{263}{145}
\parbox{3.5in}{\phantom{dummy}} \hfill
\parbox{3.5in}{\protect\caption{\vglue -3.0in 
\footnotesize
{\sc Fig}.~~7---
The {\Lya} {\NV}, {\CIV}, and {\SiIV} absorption lines of
PC~1415+3408.
For {\Lya} and {\NV}, both the 1200 and 300~lines~mm$^{-1}$ spectra
are presented so that the greater detail at higher resolution can be
compared to that at lower resolution.  For presentation, the spectra
have been normalized by the both the ``source'' and broad emission
line continua and are aligned in velocity space referenced to the quasar
emission redshift.  Long ticks above the continuum give the velocities
of the tentatively suggested components. The red members of the
doublets are shown with shorter ticks. In the case of {\Lya}, the
additional ticks show the expected location of {\SiIII} $\lambda 1206$.
\label{fig:fig7}
}}
\end{figure*}

In Figure~\ref{fig:fig7}, we present the {\Lya} and {\NV} absorption
(both the 1200 and 300~lines~mm$^{-1}$ spectra are shown), and the
{\SiIV} and {\CIV} absorption in the 300~lines~mm$^{-1}$ spectrum.
Both resolutions of the {\Lya} and {\NV} are displayed so that the
relative quality of the {\SiIV} and {\CIV} data can be judged.
The profiles are aligned in velocity with the zero point set by the
stronger member of each doublet with respect to the quasar emission
redshift.
The data are continuum normalized for presentation.

\subsection{Covering Fraction of the Absorbing Material}

The observations suggest that the absorbing gas highlighted in
Figure~\ref{fig:fig7} is outflowing material originating in the
immediate environment of the quasar.
There are several examples where such material does not fully occult 
the broad emission line region and/or of the quasar continuum source
(e.g.\cite{barlow}).
Here, we investigate the possibility that the BAL flow material
associated with PC~1415+3408 partially covers the emission source(s).

If $\tau$ is the effective optical depth of an absorbing cloud that
occults a fraction, $C_{f}$, of the source, the residual intensity,
$R$, in normalized units is,
\begin{equation}
R(\lambda) = \left\{ 1 - C_{f}(\lambda) \right\} 
     + C_{f}(\lambda) e ^{-\tau(\lambda)},
\end{equation}
where the first term on the right--hand side represents unocculted
photons and the second term represents unabsorbed photons.
For a doublet, the covering factor can be obtained as the solution to
\begin{equation}
\left\{ \frac{R_{r}-1+C_{f}}{C_{f}} \right\} ^{ \tau _{b} / \tau _{r} }
 = \left\{ \frac{R_{b}-1+C_{f}}{C_{f}} \right\} ,
\end{equation}
where $\tau_{b}/\tau_{r} = (g_{L}f_{LU}\lambda )_{b} /
(g_{L}f_{LU}\lambda )_{r}$, where $g_{L}$ is the degeneracy of the
lower level, $f_{LU}$ is the oscillator strength of the transition
(where $L$ and $U$ denote the lower and upper levels, respectively),
and the subscripts $b$ and $r$ denote the blue and red transitions,
respectively.

To apply this technique, however, the absorption profiles need to be
resolved, for if they are not resolved then smearing from the
instrumental spread function destroys the optical depth ratios at each
velocity point (see Ganguly \etal 1999\nocite{rajib}).
The {\NV} doublet is resolved, though its members are partially
blended.
Using software kindly provided by R.~Ganguly, we computed a covering
factor, $C_{f}$, of $0.97^{+0.03}_{-0.12}$, over the limited velocity
region $-1100$ to $-600$~{\kms} in the 1200 line~mm$^{-1}$ spectrum.
The {\NV} absorption is consistent with a unity covering factor.
The {\CIV} doublet (300 line~mm$^{-1}$ spectrum) is fully blended, and
is not viable for the partial covering test.

For the unresolved {\SiIV} profiles, the covering factor can be
investigated using Voigt profile (VP) decomposition, which effectively
deconvolves the instrumental profile from the optical depth model.
In the case of partial covering, VP components cannot be made
consistent with both members of the doublet.
Since a VP decomposition is a non--unique model, however, any evidence
for partial covering using this technique is less conclusive.
If a {\it reasonable\/} (i.e.\ physically motivated) VP decomposition
can be made consistent with full coverage, it can tentatively be
interpreted that the data are consistent with a unity covering factor.

Using {\sc minfit} (\cite{mythesis}), which uses Levenberg--Marquardt
least square minimization (\cite{more78}) while enforcing a minimum
number of components, a VP decomposition of the {\SiIV} doublet 
yields three components consistent with a unity covering factor.
The reduced $\chi ^{2}$ was 1.2.
These three components effectively correspond to the five
components found in the higher resolution spectrum (see
Figure~\ref{fig:fig6}), with the central, narrow {\SiIV} VP component
\pagebreak
corresponding to the central {\Lya} Gaussian component, and the VP
components in the wings of the {\SiIV} each corresponding to two
Gaussian components in the wings of the {\Lya}.

\subsection{Lower Limits on Column Densities}

\begin{deluxetable}{lccc}
\tablenum{3}
\tablecolumns{4}
\tablewidth{350pt}
\tablecaption{Absorption Line Properties of PC~$1415+3408$}
\tablehead
{
\colhead{Ion} &
\colhead{$W_{\rm r}$, {\AA}} &
\colhead{$\log N_{aod}$, {\cmsq}} &
\colhead{$(v^{-},v^{+})$, {\kms}} 
}
\startdata 
{\HI}\tablenotemark{a}         \dotfill & $3.7\pm0.3$ & $>18.2$ & $(-2700,-200)$
  \\
{\NV} 1239\tablenotemark{b}         \dotfill & $5.3\pm0.2$ & $>15.7$\tablenotemark{c}  & $(-2200,-400)$ \\
{\CIV} 1548       \dotfill & $5.5\pm0.9$ & $>15.4$\tablenotemark{c}  & $(-2300,+
100)$  \\
{\SiIV} 1394 \dotfill & $2.7\pm0.3$ & $14.7\tablenotemark{d} $ & $(-1900,-600)$ 
\\
{\SiIV} 1402 \dotfill & $2.9\pm0.3$ & $15.0$\tablenotemark{d}  & $(-1900,-600)$ 
\\
\enddata
\tablenotetext{a}{The tabulated column density is based upon Lyman limit
break in 300 line~mm$^{-1}$ spectrum (see text).  The apparent optical
depth method yielded $\log N_{aod}({\HI}) > 15.3$~{\cmsq}.}
\tablenotetext{b}{Measured from the 1200 line~mm$^{-1}$ spectrum.}
\tablenotetext{c}{Ambiguity due to doublet blending.  A unity doublet
ratio has been assumed.}
\tablenotetext{d}{These column densities should be considered lower
limits due to unresolved saturation.}
\label{tab:absorption}
\end{deluxetable}

Though BAL absorption lines sample dynamically active gas over a wide
range of ionization conditions and kinematics, the lines are often 
heavily blended so that deducing the physical conditions of the gas
can be intractable (\cite{arav98}; however, see \cite{turnshek96}).
In Table~\ref{tab:absorption}, we show the lower limits on the column
densities using the apparent optical depth (AOD) method
(\cite{savage91}).
The {\NV} and {\CIV} doublets are both blended and saturated.
For the {\SiIV}, the profiles do not drop to zero flux in their cores,
but they do exhibit unresolved saturation, as deduced by the factor of
two difference in their apparent column densities; as such, the
measured {\SiIV} column densities are likely to be significantly
underestimated.
The data do not allow definitive statements about the metallicities
and ionization conditions of the absorbing gas.

\section{Discussion}
\label{sec:summary}

The UV rest--frame absorption associated with PC~1415+3408 exhibits
four remarkable properties.

(1) The {\NV} doublet is consistent with black saturation over six
resolution elements in a 1.7~{\AA} resolution spectrum.  The doublet
is partially blended. 

(2) The {\CIV} is consistent with black saturation in a $\sim
8$~{\AA}, lower resolution spectrum.

(3) The {\Lya} absorption along the blue wing of the {\Lya} emission
line is clearly associated with the primary {\NV} and {\CIV} troughs
(velocity range $-1700 \leq v \leq 0$~{\kms}).
This {\Lya} absorption is neither smooth nor flat bottomed, but
exhibits clear structure. 

(4) Both {\CIV} and {\NV} have secondary ``detached'' troughs, and
these troughs are not aligned in velocity space; {\NV} is at lower
velocities than {\CIV}.
The detached troughs have velocity spreads less than 2000~{\kms} and
smooth, non--zero, flat bottoms.

What makes the absorption in PC~1415+3408 exceptional is that {\it
both\/} the {\NV} and {\CIV} absorption profiles are consistent with
black saturation, even at a spectral resolution of $\sim 8$~{\AA}
(300~lines~mm$^{-1}$ spectrum).
In the 1200 lines~mm$^{-1}$ spectrum (resolution 1.7~{\AA}), the
unblended portion of the {\NV} $\lambda 1238$ transition is consistent
with zero flux across six resolution elements.
It is unusual for this to occur with {\CIV} [especially in comparable
5--7~{\AA} resolution spectra (e.g.\ \cite{apmsurvey})] and it is
exceedingly rare with {\NV} at all observed resolutions.
Interestingly, the Mini--BALs 0835+5804 and PHL~1157+0128 have
\pagebreak
absorption nearly consistent with being black (\cite{aldcroft97}).
Though the velocity spreads of mini--BAL quasars are much smaller than
that of ``classical'' BAL quasars, it is of interest to make a direct
comparison between PC~1415+3408 and BAL quasars in light of a unified
model of the BAL phenomenon.
In the 72 BAL quasar spectra observed by Korista \etal
(1993\nocite{korista93}) at resolution 2.1~{\AA}, only two objects
have zero flux in the {\CIV} absorption profile.
Of the 58 Korista \etal spectra that cover {\NV} and {\Lya}, none
have black {\NV} absorption, even when the {\NV} profiles are
saturated.

Black absorption troughs imply that the absorbing material is being
viewed from a direction where both the broad emission line region and
the continuum sources are fully occulted by the BAL material and there
is virtually no scattering back into the line of sight.
Since black saturation is virtually never seen in BAL quasars, this
suggests that we are either seeing PC~1415+3408 at a low probability viewing
angle, or that the BAL flow geometry around this quasar is not common
to BAL quasars.
For BALs originating from an equatorial flow (e.g.\ \cite{emmering92};
\cite{murray95}; \cite{dekool95}), two low--probability viewing
angles include one {\it directly\/} edge on to the disk or one grazing
the BAL flow material right at the opening angle.
However, a line of sight grazing the opening angle might be more
likely to have residual flux in the absorption trough due to 
partial coverage of the compact continuum source or due to light
scattered back into the line of sight or to a non--unity filling
factor of BAL material.
It might be that the spatial extent or cross section of any scattering
material is diminutive.
On the other hand, an edge--on viewing angle might be difficult to
understand in view of the larger overall velocity spread predicted by
some models for such an orientation (e.g.\ see Fig.\ 5$b$ of
\cite{murray95}).
Whatever the viewing angle, the kinematics of the {\Lya} absorption
and the presence of {\NV} and {\CIV} detached troughs suggest that the
BAL flow material does not have uniform ionization and/or density
structure.


The velocities of the secondary, or detached troughs, are $-3500 \leq v
\leq -2200$~{\kms} for {\NV} and $-5000 \leq v \leq -3000$~{\kms} for
{\CIV} (see Fig.~\ref{fig:fig7}). 
Though they are narrower, these troughs are more typical of the
non--black, flat--bottomed, saturated profiles seen in higher velocity
BALs (e.g.\ \cite{arav98}).
Since the detached {\NV} trough is at a lower velocity than the 
detached {\CIV} trough, the ionization condition may decrease with
outflow velocity.
This is contradictory to models that predict constant ionization [see
de~Kool (1997\nocite{dekoolAGN}) and references therein] or increasing
ionization (e.g.\ \cite{murray95}) with outflow velocity.
For the {\NV}, either the continuum source is not absorbed
(occulted) or the trough is significantly filled in by photons.
The latter would imply that any material that scatters light back into
the line of sight would be at relatively high velocities with respect
to the quasar.
It should be noted that there is absorption structure along the red
wing of the {\Lya} emission line that may be due to even higher
velocity {\NV} gas (see upper panel of Figure~\ref{fig:fig6}).
In the case of {\CIV}, at least part of the continuum source is
absorbed, or the source of photons filling in the absorption trough
is not as strong as for the {\NV}.

Double--trough broad absorption from {\CIV} has been investigated as a
possible signature of line--driven radiation pressure.
Since all absorbing species experience the same flow, they each have
the same optical depth modulation, showing a reduced optical depth due
to increased flux from the {\Lya} emission line.
The so--called ``ghost of {\Lya}'' effect (\cite{turnsheketal88};
\cite{aravAGN}) corresponds to a ``hump'' in the {\CIV} absorption at
$ v \sim -5900$~{\kms}, the velocity of the {\Lya} emission peak in
the rest--frame of {\NV} ions.
Given the exceptionally strong {\NV} absorption, PC~1415+3408 is a
good candidate for exhibiting this modulation if the material is
driven by radiation pressure.
The ghost of {\Lya} effect is not seen in PC~1415+3408; however, we
point out that the {\CIV} absorption terminates at $\sim -5000$~{\kms}
and that this termination could be induced by the effect.

\section{The Promise of Mini--BAL Quasars}

Despite significant efforts, the overall multi--band statistical
properties of quasars exhibiting BAL flows are not yet on a solid
statistical footing.
Overall, BAL quasars are radio--quiet (e.g.\ \cite{stocke92}) and
quiet, or self--absorbed, in soft X--rays (\cite{kopko94};
\cite{green96}).
However, the FIRST Survey has produced examples of radio--selected,
radio--loud BAL quasars (\cite{becker97}), and there are now a few
X--ray loud BAL quasars observed in hard X--rays (e.g.\
\cite{mathur95}; \cite{gallagher}; and see predictions of
\cite{krolik98}).
For radio--loud BAL quasars, the radio spectra have a range of
spectral indices (e.g.\ \cite{barthelAGN}; \cite{becker97}), and it is
not unexpected that X--ray spectra would as well.

The X--ray spectra of Mini--BAL quasars may be a key for understanding
BAL flows.
For example, assuming radiation pressure driven flow dynamics, Murray
\etal (1995\nocite{murray95}) predict that quasars with harder
(flatter) X--ray spectra, will have wind terminal velocities no
greater than $\sim 5000$~{\kms}.
Flatter X--ray spectra might naturally explain the sub--population of
Mini--BAL quasars, if the flow dynamics are governed by the energy
index, $\alpha_{E}$, of the X--ray spectrum.
Mini--BALs may provide the first robust measurements of or limits on
the X--ray spectral energy indices, especially using more powerful,
forthcoming X--ray observatories ({\it AXAF\/} and {\it XMM\/}). 
Both deeper radio and hard X--ray observations of Mini--BAL quasars 
at all redshifts could be very telling for constraining unified models
of quasars.
At the highest redshifts, PC~1415+3408 appears to be a good target for
for such investigations.

\acknowledgements
We especially thank Niel Brandt for many fruitful discussions on BAL
quasar and X--ray properties and David Turnshek for comments that led
to a much improved manuscript.
We also thank Roger Blandford, Jane Charlton, George Chartas, Mike
Eracleous, and Sarah Gallagher for helpful discussions, and Rajib
Ganguly for use of his code to compute partial covering factors.
Robert Deverill and  David Saxe created much of the software for the
data processing and George Weaver provided valuable computing support.
This work has been supported in part by National Science
Foundation Grants AST96--17185 (CWC), AST95--09919 (DPS), AST94--15574
(MS), AST86--18257A02 (JEG), and by NASA LTSA Grant NAG5--6399 (CWC).

\end{document}